\documentstyle[12pt]{article}

\newcommand{\beqn}{\begin{equation}}
\newcommand{\eeqn}{\end{equation}}
\newcommand{\beqa}{\begin{eqnarray}}
\newcommand{\eeqa}{\end{eqnarray}}
\newcommand{\beqas}{\begin{eqnarray*}}
\newcommand{\eeqas}{\end{eqnarray*}}
\newcommand{\n}{\nonumber}
\newcommand{\ep}{\epsilon}

\begin{document}
\begin{center}
{\bf Loop Variables in Topological Gravity}\\
\vspace{15 pt}
{\it by}\\
\vspace{13 pt}
Y. Bi {\it and}  J. Gegenberg\\[5pt]
{\it Department of Mathematics and Statistics}\\
   {\it University of New Brunswick}\\
   {\it Fredericton, New Brunswick}\\
   {\it CANADA E3B 5A3}\\[5pt]
\end{center}
\vspace{20pt}
{\narrower\smallskip\noindent
{\bf Abstract}:  We examine the relationship between covariant and
canonical (Ashtekar/Rovelli/Smolin) loop variables in the context
of BF type topological field theories in 2+1 and 3+1 dimensions,
with respective gauge groups SO(2,1) and SO(3,1).  The latter model
can be considered as the simplest topological gravity theory in 3+1
dimensions.  We carry out the canonical quantization of this model
in both the connection and loop representations, for the two
spatial topologies $T^3$ and $S^2\times S^1$. \\
\vspace{10 pt}
\begin{center}March 1993
\end{center}
UNB Technical Report 93-02
\newpage
\section{Introduction}
An important advance in the quest to construct a viable and
realistic
quantum theory of gravity was made by Ashtekar and his coworkers
when
they recast and slightly generalized Einstein gravity by
introducing
new canonical variables \cite{ash:bo,jac:nul}.  In these new
variables, the constraints are polynomial and can be solved
explicitely in terms of non-local, but gauge invariant, "loop
variables" $T^n$ \cite{jac:nul,rov:nul}.  These loop variables are
not the
{\it physical} observables in 4-D gravity since they are not
diffeomorphism
invariant \cite{pullin}.  The discovery of non-trivial physical
observables
is an outstanding problem in the non-perturbative
quantization of 4-D Einstein gravity.
\par In three dimensional gravity theory the problem of
constructing physical
observables is at least partially solved.  The reason for this is,
essentially, that three
dimensional gravity
is a topological field theory \cite{wt:nul}.  Indeed, if we relax
the
non-degeneracy of the spacetime triad, in the case of zero
cosmological
constant, 3-D Einstein gravity (in the first order formalism) is
equivalent to
Chern-Simons theory with gauge group $ISO(2,1)$.  In fact, Einstein
gravity
in the first order formalism is the prototype of another
interesting
topological field theory- the BF theories \cite{ho:com,bl:pl}.
With non-zero
cosmological constant, Einstein gravity is equivalent to a
Chern-Simons
theory with gauge groups SO(3,1) and SO(2,2), respectively, in the
cases of positive and negative cosmological constant.  The
equations of
motion (or constraints) imply that the $ISO(2,1)$ connection in the
Chern-Simons model is flat, i.e., locally pure gauge.  One may now
construct
gauge invariant observables that do not vanish when the
constraints are imposed and which are diffeomorphism invariant
(due to the fact that the diffeomorphisms are generated by the
ISO(2,1)
gauge transformations in this case \cite{wt:nul}).  These are {\it
physical} observables, denoted generically here by $W$, the
``covariant loop variables'' of the system.
The canonical quantization of 3-D Einstein gravity in first order
formalism
can be handled very much analogously
to the 4-D case \cite{smo:hop}.  In particular, the constraints can
be solved
in terms of the 3-D analogues of the $T^n$, i.e., "Ashtekar" loop
variables.
\par
In the next section we will elucidate the relationship between
 the two species of global observables encountered
in 3-D gravity theory: the covariant loop variables and the
Ashtekar
loop variables.  In section 3. we will review the properties of 4-D
BF
theory with gauge group SO(3,1).  This theory may be viewed as the
closest
topological field theory relative of 4-D Einstein gravity
\cite{ho:com,japan}
.  We will canonically quantize this theory in terms of the
Ashtekar loop
variables $T^n$ and relate the latter to the covariant loop
variables of
the theory.  Finally, in section 4. we will first of all use the
results obtained to explicitely carry out the quantization of this
model for the special cases  where 3-space has the topologies $T^3$
and $S^2\times S^1$, and second we will discuss the implications for
non-topolocical quantum gravity in 4-D.

\section{3-D Gravity}
We begin by reviewing 3-D Einstein gravity in the form considered
by Witten
\cite{wt:nul}.
In (2+1)-dimensions, the Einstein-Hilbert action can be written in
the form \cite{wt:nul}
\beqn
\label{EH}
S=\frac{1}{2}\int
d^3x\epsilon^{ijk}\epsilon_{abc}e^{a}_{i}R^{bc}_{jk},
\eeqn
where lower case latin indices from the beginning of
the alphabet are $SO(2,1)$ indices, while those from the middle
are spacetime indices.
The $e^{a}_{i}$
are an orthonormal triad and $\omega^{ab}_{i}$ are the spin
connection
components with respect to the group $SO(2,1)$. The curvature is
given
by
\begin{equation}
R^{ab}_{ij}=
\partial_{i}\omega^{ab}_{j}-\partial_{j}\omega^{ab}_{i}+
             [\omega_{i}, \omega_{j}]^{ab},
\end{equation}
and the classical equations of motion are
\begin{eqnarray}
D_{i}e^{a}_{j}-D_{j}e^{a}_{i}=0,                 \\
R^{ab}_{ij}=0.
\end{eqnarray}
Here $D_{i}$ is the covariant exterior derivative with respect to
the
connection $\omega$.  The equations of motion imply that the
spacetime
is a flat Lorentzian manifold. The gauge group of the system is the
Poincar\'{e} group $ISO(2,1)$.
Let $J^{ab}$ be Lorentz generators and $P^{a}$ the translations, so
that the
gauge potential for the group $ISO(2,1)$ is the one-form
\begin{equation}
\label{con}
A_{i}=e^{a}_{i}P_{a}+\omega^{a}_{i}J_{a}.
\end{equation}
Here we have defined
$\omega^{a}_{i}:=1/2\epsilon^{abc}\omega_{ibc}$ and
$J_{a}:=1/2\epsilon_{abc}J^{bc}$. The generators $P_{a}$, $J_{a}$
determine
a quadratic form on the Lie algebra
\begin{equation}
\label{qut}
<J_{a},J_{b}>=<P_{a},P_{b}>=0,\,\,\, <J_{a},P^{b}>=\delta_{a}^{b}.
\end{equation}
The generators of $ISO(2,1)$, $P_{a}$ and $J_{a}$, satisfy the
algebra
\begin{eqnarray}
\label{alg}
[J_{a},P_{b}]=\epsilon_{abc}P^{c},   \\
 {[J_{a},J_{b}]}=\epsilon_{abc}J^{c},   \\
  {[P_{a},P_{b}]}=0.
\end{eqnarray}
The Chern-Simons functional for the $ISO(2,1)$ gauge connection
$A_{i}$,
\begin{equation}
I_{CS}=\int_{M}Tr(A\wedge dA+\frac{2}{3}A\wedge A\wedge A),
\end{equation}
precisely coincides with the Einstein-Hilbert action (\ref{EH}), as
follows from equations (\ref{con})--(\ref{alg}).
Hence (2+1)-dimensional gravity is a Chern-Simons gauge field
theory.

We perform a canonical analysis in order to display the
constraints.
Let $M=R\times \Sigma$, where $\Sigma$ is a closed 2-manifold. The
action can
be written as
\beqa
\label{ham}
S=\int
dt\int_{\Sigma}d^{2}x{(\epsilon^{abc}\dot{\omega}^{\mu}_{bc}}
\epsilon_{\mu\nu}e_{a}^{\nu}             \n            \\
+{\frac{1}{2}\epsilon^{abc}\omega_{0bc}\epsilon^{\mu\nu}D_\mu
e_{\nu a}}
+{\frac{1}{2}\epsilon^{abc}\epsilon^{\mu\nu}e_{0a}R^{bc}_{\mu\nu}
)},
\eeqa
where the $\epsilon^{\mu\nu}:=\epsilon^{0\mu\nu}$, and the dot
``$\cdot$''
denotes differentiation with respect to $x^0$. The 0-components of
the
fields are Lagrange multiplers, which impose the constraints
\beqn
\begin{array}{l}
D_{[\mu}e^{a}_{\nu]}\approx 0,    \\
R^{ab}_{\mu\nu}\approx 0,
\end{array}
\label{cons}
\eeqn
where $\mu,\;\nu=1,\,2$ are spatial indices. Note that they are of
the same
form as the equations of motion of the gauge fields $(e,\omega)$.
{}From the
action (\ref{ham}), the fundamental Poisson bracket can be read off
as
\beqn
\{\omega_{\mu}^{a}(x), \tilde{e}^{\nu}_b(y)\}=
\delta_{b}^a\delta_{\mu}^{\nu} \delta^2 (x,y),
\eeqn
where $\tilde{e}^{\nu}_b=\epsilon^{\nu\rho} e_{\rho b}$ is a
density.

In 2+1 gravity, the covariant gauge invariant observables are the
Wilson loop
variables of the connection
$A_{i}=e^{a}_{i}P_{a}+\omega^{a}_{i}J_{a}$:
\beqn
W_R
(C)=Tr_{R}P\exp\{\oint_{C}(e^{a}_{i}P_{a}+\omega^{a}_{i}J_{a})dx^
{i}\},
\eeqn
i.e., the $ISO(2,1)$ holonomy operators \cite{wt:nul}. The
spacetime curve $C$
is closed and $R$ is a representation of $ISO(2,1)$.
\par
Variables analogous to
those introduced by Ashtekar in the 3+1 dimensional case \cite{ash:bo}
can be
constructed in 2+1-D theory \cite{smo:hop}. From the Hamiltonian
form of the
Einstein-Hilbert action, a class of loop variables on the conjugate
variables
$(e, \omega)$ can be defined. The first of these observables is the
{\it spatial} Wilson loop of \(\omega\) ( the $SO(2,1)$
connection):
\beqn
T_r^0 [C]:=Tr_r  P\exp[\oint_C{\omega}]=
Tr_r P\exp[\oint_C{ds^{\mu} \omega^a_\mu J_a}],
\eeqn
where $r$ denotes a representation of $SO(2,1)$ and $C$ is a loop
in space.
The remaining variables, denoted $T^n_r [C]$, depend on the
$e$-variables.
They are obtained by inserting $n\;\,e$-variables along the
holonomy of a loop:
\beqa
T^n_r [C]=\int ds^{\mu_1}\cdots \int ds^{\mu_n}{\sum_p{\theta
(s_q-s_p)
\cdots \theta (s_j-s_i)}}\times     \n       \\
        {{Tr_{r} [e_{\mu_i}(C(s_i))U_C (s_i,
s_j)e_{\mu_j}(C(s_j))\cdots
         e_{\mu_p}(C(s_p))U_C (s_p, s_i)]}},
\eeqa
where the  summation is over the permutation of $n$ indices and
$\theta(s_i-s_j)$ are step functions.
In the following, we omit the
representation subscript $r$. The quantity $U_C$ is defined as
\[
U_C (s,t):=P\exp[\int_s^t{ds^\mu \omega^a_\mu J_a}]\,.
\]
The $T^n$'s form a closed graded Poisson algebra, which has the
structure
\beqn
\{T^n, T^m\}\sim T^{n+m-1}.
\eeqn
{}From this we can see that $T^0$ and $T^1$ form a subalgebra.
In the case of three dimensional gravity, we need only consider the
$T^0,\;T^1$
variables. From Ref.\cite{smo:hop} we know that the Poisson
brackets of the
observables with the constraints are weakly zero. Hence $T^0, T^1$
are physical
observables for pure  2+1 gravity. When gravity is coupled to
ordinary local
matter, then $T^0, T^1$ are no longer the physical observables
because they
fail to commute with all the  constraints. However they are still
gauge
invariant and very important quantities for the construction of the
loop
representation \cite{ash:bo,kim:cqg}.

Now we have two kinds of observables-- Ashtekar $T$ variables and
the
covariant loop variables $W$-- for 2+1 gravity theory. We now
discuss the relation
between these two kinds of observables in the canonical formulation
of the
theory. The loops $C$ are taken to be at fixed time, i.e. they lie
in
a two-dimensional spatial hypersurface $\Sigma_t$. So in the
covariant loop
variables
\beqn
W(C)=TrP\exp\{\oint_{C}(e^{a}_{\mu}P_{a}+\omega^{a}_{\mu}J_{a})dx
^{\mu}\},
\eeqn
the curves $C$ are spatial, as are the loops in the Ashtekar $T$
variables.
Since the $ISO(2,1)$ group manifold is the total space of the
cotangent
bundle of the $SO(2,1)$ group manifold \cite{wt:nul}, we can
express the
translation generators
$P_\mu$ in terms of the rotation generators $J_\mu$ and an
infinitesimal
parameter $\theta$ \cite{mar:nuc}:
\beqn
\label{P}
P_\mu=\theta J_\mu,
\eeqn
where $\theta$ is taken to satisfy $\theta^2=0$ and its trace is
given as
$tr\theta=c$, where $c$ is a constant. Substituting  (\ref{P}) into
$W(C)$,
we have
\[
W(C)=TrP\exp\{\oint_{C}(\theta e^{a}_{\mu}J_{a}
+\omega^{a}_{\mu}J_{a})dx^{\mu}\}.
\]
Within the path order operator, all the variables commute. So we
can rewrite $W(C)$ as
\beqa
W(C)=TrP[\exp{\oint_{C}\theta e^{a}_{\mu}J_{a}dx^\mu}
exp{\oint_{C}\omega^{a}_{\mu}J_{a}dx^{\mu}}]   \n \\
=TrP[\sum_{n=0}^{\infty} \frac{\theta^n}{n!}{(\oint_{C}
e^{a}_{\mu}J_{a}dx^\mu)}^n
\exp{\oint_{C}\omega^{a}_{\mu}J_{a}dx^{\mu}}].
\eeqa
{}From the properties of $\theta$, we know that $\theta^n=0$ for
any $n\geq 2$. Therefore
\beqa
\label{rel}
W(C)&=&TrP\exp[\oint_C \omega^a J_a]
+Tr\theta P(\oint dx^\mu e^a_\mu J_a\exp[\oint_C \omega^a J_a]) \n
\\
&=&TrP\exp[\oint_C \omega^a J_a]
+c Tr\oint dx^\mu e^a_\mu (C(s))U_C(s)     \n \\
&=&T^0[C]+cT^1[C].
\eeqa
We conclude that the covariant loop variables for spatial loops are determined
by the Ashtekar $T$ variables.

If point particles are present, we locate them at punctures in
space
\cite{mar:nuc}.  We compute these two kinds of  observables for
loops $C$ which
enclose only one puncture. In order to evaluate the observables, we
first
choose the representation of $ISO(2,1)$ given by
\beqn
\label{rpt}
  J_0=\frac{1}{2} \left( \begin{array}{cc}
                          i &0\\
                          0 &-i
                         \end{array} \right)  \;\;,
  J_1=\frac{1}{2} \left( \begin{array}{cc}
                          0 &i\\
                         -i &0
                         \end{array} \right)  \;\;,
  J_2=\frac{1}{2} \left( \begin{array}{cc}
                          0 &1\\
                          1 &0
                         \end{array} \right)  \;\; ,
\eeqn
\beqn
P_a=\theta J_a.
\eeqn
In this representation we have \cite{mar:nuc}
\beqn
\label{pro}
J_aJ_b=-\frac{1}{4}\eta_{ab}+\frac{1}{2}\epsilon_{abc}J^c\; ,
\eeqn
Using the fundamental Poisson brackets and this representation of
$SO(2,1)$,
we compute the loop algebra for $T^0, T^1$:
\beqn
\{T^0 [\gamma], T^0 [\delta]\}=0,
\eeqn
since the $T^0$ just depend on connection $\omega$, and
\beqa
\{T^1 [\gamma], T^0 [\beta] \}=
\sum_i \Delta_i (\gamma, \beta)(T^0[\gamma\#_i
\beta]-T^0[\gamma\#_i \beta^{-1}])     \\
\{T^1 [\gamma], T^1 [\beta] \}=
\sum_i \Delta_i (\gamma, \beta)(T^1[\gamma\#_i
\beta]-T^1[\gamma\#_i \beta^{-1}]),
\eeqa
where $i$ labels the points where the loops $\gamma$ and $\beta$
intesect,
$\#_i$ stands for composition of two loops at the $i^{th}$
intersection.
The structure constants of the loop algebra are
\beqn
\Delta_i (\alpha, \beta):=\int_i {ds^\mu dt^\nu \delta^2(\alpha
(s), \beta(t))}
                          {\epsilon_{\mu\nu}},
\eeqn
where the integrals are taken in an interval including the $i^{th}$
intersection.  The number $\Delta_i$ is equal to minus or plus one depending on
whether
the dyad formed by the tangent vectors of the two loops at the
$i^{th}$
intersection is left-- or right--handed with respect to the
orientation
given by $\epsilon_{\mu\nu}$.

The particle associated with each puncture carries mass and angular
momentum
\cite{mar:nuc}. So at the puncture the constraints (\ref{cons}) are
not
enforced.  Because of the constraints (\ref{cons}) we can make the
loop $C$
infinitesimally close to the puncture in the covariant loop
variable $W$. Using the
infinitesimal condition, we can replace the path ordered line
integral of the
connection by the area integral of the curvature enclosed
\cite{mar:nuc}, i.e.,
\beqn
W(C)=Tr\exp\{\theta j^{a}J_{a}+p^{a}J_{a}\},
\eeqn
where
\beqn
\label{ang}
p^a:=\int_{\delta \sigma}R^a ,\;\; j^a:=\int_{\delta \sigma}De^a,
\eeqn
are the area integrals of the curvatures concentrated at the
puncture. From the
formula
\beqn
\label{gro}
\exp(N \cdot J)=\cos(|N|/2)+2\sin(|N|/2)\frac{N\cdot J}{|N|},
\eeqn
and the tracelessness of $J_a$ , we compute the covariant loop
observables
to be
\[
W(C)=Tr \cos(|p+\theta j|/2) .
\]
The normal of the vector is defined by
\[
|p+\theta j|:=(p+\theta j)^a(p+\theta j)_a
=|p|+\theta p\cdot j/|p|
\]
where we have used the fact $\theta^2 =0$. Again, applying this and
expanding,
we find that
\[
\cos[(|p|+\theta p \cdot j)/2]
=\cos(|p|/2)-\theta p \cdot j\frac{\sin(|p|/2)}{2|p|}.
\]
Finally the covariant loop observables for one puncture are:
\beqa
W(C)=Tr[\cos(|p|/2)-\theta p\cdot j\frac{\sin(|p|/2)}{2|p|}] \n \\
=2\cos(|p|/2)-c p\cdot j\frac{\sin(|p|/2)}{|p|}.
\eeqa
We get the same result by evaluating the $T$-variables and using
the relation
between the latter and covariant loops in (\ref{rel}).

We now briefly consider the quantum theory.  We quantize in the
reduced phase, as in Refs. \cite{ash:bo,smo:hop}, for the special
case of $\Sigma=T^2$, the flat 2-torus.  We summarize the results
here. Due to the non-connectedness of the group
$SO(2,1)$ and the Abelian property of the homotopy group of the
2-torus, the reduced
phase space has disconnected sectors. For the time-like case, the
reduced
configuration space is topologically a 2-torus. Let us choose
$a,\;b$ as its
coordinates, $a,\;b\in[0,\;1]$.  The loop variable are promoted to
operators $\hat T^0$ and $\hat T^1$ on the Hilbert space of $L^2$
functions $\Psi(a,b)$ over $\hat{\cal C}$ such that:
\beqa
\hat{T}^0[\alpha] \Psi(a,b)=2\cos(\pi a)\Psi(a,b),\\
\hat{T}^1[\alpha] \Psi(a,b)
=-2\pi i \hbar \sin(\pi a)\frac{\partial}{\partial b}\Psi(a,b).
\eeqa
It is easy to check that the {\it commutators} of these operators
are given by $-i\hbar$ times the value of the corresponding
classical Poisson brackets.
Hence the covariant loop variables on the 2-torus are given by the
relation (\ref{rel}) as
\beqn
W(\alpha)\Psi(a,b)=\{2\cos(\pi a)-2c\pi i\hbar \sin(\pi a)
\frac{\partial}{\partial b}\}\Psi(a,b).
\eeqn
\section{Observables for 4-D BF theory}
A careful analysis of Witten's approach to 2+1 dimensional gravity
reveals
that the crucial reason for the solvability of the theory is that
gravity in
three dimensions has no {\it local} dynamics. The extension of
 this model to higher
dimensions results in the BF theory
with the following form of action \cite{ho:com,bl:pl}:
\[
S=Tr\int B\wedge F \label{bft},
\]
where $F$ is the curvature two-form of a connection $A$ on a
principle
bundle over the $n$-dimensional manifold $M$ with (simple)
structure group
$G$. $B$ is a Lie algebra-valued $(n-2)$-form on $M$. On the
4-manifold
$M=R\times \Sigma$ with structure group $G=SO(3,1)$, the action can
be written
as
\beqn
S=\int P_{ab}\wedge R^{ab},
\eeqn
The action is put into the canonical form, with the result that the
phase space variables are the $\omega^{ab}_i$ and their conjugate momenta
$\pi^i_{ab}$, where
\beqn
\pi^i_{ab}=\ep^{ijk}P_{abjk}.
\eeqn
Here $a,\,b,...$ are SO(3,1) indices and $i,\, j,...$ are the
spatial indices
on $\Sigma$. The corresponding loop variables are, first, the
$T^0$,
given by
\beqn
T^0[\gamma]=Tr U_{\gamma}(s),
\eeqn
where
\beqn
U_{\gamma}(s)=P\exp(\oint_{\gamma} \omega^{ab}J_{ab}),
\eeqn
with $\gamma$ a closed loop on $\Sigma$, and $J_{ab}$ the
generators
of $SO(3,1)$; and second the $T^i$, given by
\beqn
T^i[\gamma](s)=Tr[\pi^i(\gamma(s))U_{\gamma}(s)],
\eeqn
which are linear in the momenta. Applying the fundamental Poisson
bracket
\beqn
\label{fund}
\{\omega_{i}^{ab}(x), \pi_{cd}^{j}(y)\}
=\delta_{i}^{j}\delta_{c}^{[a}\delta_{d}^{b]}\delta(x,y),
   \\
\eeqn
we can compute the loop algebra. Obviously, we have
\beqn
\{T^0[\gamma],\,T^0[\beta]\}=0.
\eeqn
The important Poisson bracket is between $T^0$ and $T^i$.  First
\beqas
\{T^0[\gamma],\, T^i[\beta](t)\}=
\{{U_{\gamma}(s)^I}_{\mbox{}I},\,
{\pi^i(\beta(t))^K}_{\mbox{}L}{U_\beta(t)^L}_{\mbox{}K}\}        \\
=\int dx^i \delta(\beta(t),\,
\gamma){U(s,t)^I}_{\mbox{}J}J_{ab}^{JM}
U(t,s)_{MI} {J^{abK}}_{\mbox{}L}{U_\beta(t)^L}_{\mbox{}K} ,
\eeqas
where (\ref{fund}) has been used. If we choose the representation
of
generators $J_{ab}$ in $4\times 4$ form\cite{tun:ki}, we have
\[
{{J_{ab}}^J}_{\mbox{}M}=\delta^J_a\eta_{bM}-\delta^J_b\eta_{aM}.
\]
It follows that
\[
{J_{ab}}^{JM}{J^{abK}}_L=2(\eta^{JK}\delta^M_L-\eta^{MK}\delta^J_
L).
\]
Using this result and noticing that $U_\gamma (s)_{IJ}$ is an
element of
$SO(3,1)$, and satisfies the identity
\footnote{We use the correspondence between Lorentz tensors and
SL(2,C) spinors to write $U_\gamma(s)_{IJ}$ as follows:
\[U_\gamma
(s)_{IJ}=\sigma_I^{aa^{\prime}}\sigma_J^{bb^{\prime}}U_\gamma(s)_
{ab}
\bar{U}_\gamma (s)_{a^{\prime}b^{\prime}},\]
where the matrices $U_\gamma(s)_{ab}$ and its conjugate are the
``SL(2,C)'' representations of the holonomy operator along the
curve
$\gamma(s)$.  It is shown in \cite{ash:bo} that these satisfy:
\[U_\gamma (s)_{ab}=-U_{\gamma^{-1}} (s)_{ba};\;\;
\bar{U}_\gamma (s)_{a^{\prime}b^{\prime}}=
-\bar{U}_{\gamma^{-1}}(s)_{b^\prime a^\prime}.\]
Hence
\[U_\gamma
(s)_{IJ}=\sigma_I^{aa^{\prime}}\sigma_J^{bb^{\prime}}U_{\gamma^{-
1}}(s)_{ba}
\bar{U}_{\gamma^{-1}}
(s)_{b^{\prime}a^{\prime}}=U_{\gamma^{-1}}(s)_{JI}.\]}
\[ U_\gamma (s)_{IJ}=U_{\gamma^{-1}} (s)_{JI},\]
we obtain
\beqn
\{T^0[\gamma],\, T^i[\beta](t)\}
=2\int dx^i \delta^3(\beta(t),\, \gamma)
[T^0(\gamma \# \beta)-T^0(\gamma\# \beta^{-1})].
\label{fac}
\eeqn
This expression contains distributional factors
\[
2\int dx^i \delta^3(\beta(t),\, \gamma),
\]
as does the local expression(\ref{fund}), but with the difference
that the distributional factors in (\ref{fac}) have support on
curves
rather than points.  We may replace the $T^i$ by smeared loop
variables in order to eliminate the distributional factors
in the Poisson bracket.  An elegant discussion is given by Smolin
\cite{smo:lec}, and we merely recapitulate it here.
 Since the distributional factors in
(\ref{fac}) are already one dimensional, the integral over the test
functions should be two dimensional. It is natural to make this an
integral over a surface.  Hence we consider the dual of the
momentum
$\pi^i_{ab}$ on $\Sigma_t$, defined as
\beqn
\pi^{*}_{ijab}=\frac{1}{2}\ep_{ijk}\pi^k_{ab}.
\eeqn
We note here that the $\pi^{*}_{ijab}$ are the spatial projections of the
components of the
original variables $P_{ab}$.  Correspondingly we have
\beqn
T^{*}_{ij}[\gamma](s)=Tr[\pi^{*}_{ij}(\gamma(s))U_{\gamma}(s)].
\eeqn
Let us consider a one-parameter continuous family of loops
$\gamma^i (s,u)$
where $u\in [0,1]$ such that they form a strip, which we denote
with a hat
$\hat{\gamma}$. For each $u$, \(\gamma^i (s)_u\equiv
\hat{\gamma}^i(s,u)\) is a
closed loop.  The parameters $s$ and $u$ then coordinatize the two
dimensional
surface of the strip. The following smeared version of the $T^1$
observable can
be defined\footnote{The normalization here differs by the factor
$\pi^{-2}$
 from the literature, e.g. Ref.[8].  It is necessary to choose this
 normalization in order to get the expression (69) in the form given
in the literature without clumsy factors of $\pi$.}:
\beqn
T^1[\hat{\gamma}]:={1\over\pi^2}\int du \int ds \frac{\partial
\hat{\gamma^i}} {\partial u}\frac{\partial \hat{\gamma^j}}{\partial
s} T^{*}_{ij}[\gamma](s).  \eeqn
The Poisson bracket of this
observable with $T^0[\alpha]$ is expressed in terms of the
intersection number of loop $\alpha$ and the strip $\hat{\gamma}$. If
the loop does not intersect the strip, the Poisson bracket is zero.
If the loop intersects the strip, at intersection $n$, we have \beqn
I_n (\hat{\gamma},\alpha)=\int_{\sigma_n} dS^{ij}\int dx^k \ep_{ijk}
\delta^3 (\hat{\gamma}, \alpha), \eeqn which is equal to $\pm 1$
according to the orientation of the intersection.  In the above
$\sigma_n$ denotes the neighbourhood of the intersection $n$.
Finally, the loop algebra becomes \beqa \{T^0[\alpha],\,
T^1[\hat{\gamma}](t)\}=\sum_n I_n(\hat{\gamma},\alpha)
(T^0[\alpha\#_n \hat{\gamma}]-(T^0[\alpha\#_n \hat{\gamma}^{-1}]),\\
\{T^1[\hat{\alpha]},\, T^1[\hat{\gamma}](t)\}=\sum_n
I_n(\hat{\gamma},\hat {\alpha}) (T^1[\hat{\alpha}\#_n
\hat{\gamma}]-(T^1[\hat{\alpha}\#_n \hat{\gamma}^{-1}]).  \eeqa The
geometrical significance of the loop algebra is that it depends on
the intersection number of loops or strips with other loops or
strips.  \par We know that $T$-variables are not the physical
observables in 3+1 Einstein gravity since they are not diffeomorphism
invariant.  However, the model that we consider here, unlike Einstein
gravity, is a topological field theory.  For the BF theory considered
here, it turns out that $T^0$ and $T^1$ are physical observables,
i.e.  they commute with all the constraints.  The constraints of the
4-D BF theory are
\beqa
R^{ab}\approx 0,      \\ C_{ab}:= D_i \pi
^i_{ab}\approx 0.
\eeqa
Note that the constraints $C_{ab}$ satisfy
\beqn \{C_{ab}(x),
C_{cd}(y)\}=\frac{1}{2}[\eta_{cb}C_{ad}+\eta_{bd}C_{ca}+
\eta_{ac}C_{db}+\eta_{ad}C_{bc}]\delta(x,y).
\eeqn
This is the
algebra of $SO(3,1)$, so the $C_{ab}$ are the generators of the gauge
group $SO(3,1)$. The loop variables $T^0$ and $T^1$ are gauge
invariant.  Thus it follows naturally that
\beqa
\{ T^0 [\gamma],
D_i\pi^i_{ab}\}=0     ,    \\ \{ T^1 [\hat{\gamma}], D_i
\pi^i_{ab}\}=0     .
\eeqa
Since the $T^0$ are just functions of the
configuration variables, as are $R^{ab}$, it follows trivially that
\beqn \{ T^0 [\gamma], R^{ab}\}=0 .  \eeqn The crucial Poisson
bracket is that between $T^1$ and $R^{ab}$.  Analogous to the 2+1
dimensional theory, we have
\beqn
\{T^1 [\hat{\gamma}],\,R^{ab}\}=
{1\over\pi^2}\int_\sigma D[D(\delta^{ab}_{cd}\delta^3
(\gamma(s),x)TrJ^{cd}U_\gamma(s))] \sim  R^{ab},
\eeqn
where $\sigma$
is the volume enclosed by the strip $\hat{\gamma}$ in $\Sigma$, and
we have used Stokes theorem \cite{stcs}.  This bracket weakly
vanishes by means of the constraint. This is the  desired result.

For this topological model, the reduced phase space is also a
moduli
space of flat connections {\it modulo} SO(3,1) gauge invariance,
as in 2+1 gravity. It is
interesting to consider the observables on a topologically
nontrivial underlying spatial
3-manifold $\Sigma$ which has the topology of a 3-torus $T^3$
\cite{abh:cla}.
First it is useful to discuss some features of the holonomy.
If, as we assume, reduced phase space can be polarized, then it is
the
tangent bundle over the reduced configuration space $\hat{\cal C}$.
Each element $\omega$ of $\hat{\cal C}$ is
determined by fixing a base point $p$ on $\Sigma$ and specifying
the
holonomies, {\it modulo} the action of $SO(3,1)$ at $p$, of
$\omega$ around
the $n$ generators of the homotopy group of $\Sigma$. The
holonomies
provide us $n$ gauge group elements, $(U_1,...U_n)$, and $\omega$
is
determined by the equivalence class $U\cdot (U_1,...U_n)\cdot
U^{-1}$
where $U$ is the gauge group action at $p$. $\hat{\cal C}$ has
several
disconnected components essentially because the Lie algebra
$SO(3,1)$
has three disjoint orbits under the natural action of the group.
Each
$U_k$ is a rotation either along a time-like, null, or space-like
axis
and the action of the gauge group at $p$ must map that $U_k$ to a
rotation
with the same type of axis.  For the case  that $\Sigma=T^3$, the flat
3-torus, we
have three generators for the homotopy group which are denoted as
$\alpha_1,\,\alpha_2,\, \alpha_3$. Now, since the homotopy group of
$T^3$
is Abelian, it follows that the holonomy group of any flat
connection
must also be Abelian. A general element of the homotopy group of
the
torus may be written as
\beqn
\alpha=\alpha_1^{n_1}\alpha_2^{n_2}\alpha_3^{n_3}
\eeqn
where $n_1,\,n_2,\, n_3$ are integers which represent the number of
times the
loop winds the three generators.  The holonomies satisfy
\beqn
\label{cmt}
[U_{\alpha_1},\,U_{\alpha_2}]=[U_{\alpha_3}\,U_{\alpha_2}]=
[U_{\alpha_1},U_{\alpha_3}]=0.
\eeqn
Therefore they are $SO(3,1)$ rotations around the
same axis. Under gauge transformations the axis itself rotates
preserving only
its time-like, null, or space-like character. This immediately
divides
$\hat{\cal C}$ into three sectors. In the time-like case, the
subgroup of
$SO(3,1)$ is the group of 3-dimensional rotations $SO(3)$
\cite{tun:ki} which
is compact. Thus  this sector has the topology $T^3$. In the null
sector, the
corresponding subgroup is the Euclidian group in 2-dimensions,
$E_2$. Finally
in the space-like case, the subgroup of $SO(3,1)$ is the
3-dimensional Lorentz
group $SO(2,1)$. In the cases of null and spacelike sectors, the
topologies are
more complicated.  By equation (\ref{cmt}), we should consider the
Abelian
subgroup in those sectors. Since the groups $SO(2,1)$ and $E_2$
contain
different Abelian subgroups, they may correspond to different
topologies.

We now discuss the reduced phase quantization for the time-like
case in detail. We
parameterize an arbitrary flat connection as follows:
\[ \omega=(\sum_{j=1}^3 a_j d\theta_j)\frac{i\tau^3}{2}, \]
for some choice of the real constant $a_j$, where $\theta_j$ are the
three angular
coordinates on $\Sigma$ and $\frac{i\tau^3}{2}$ one of generators
of $SO(3)$
with
\[
\label{rep}
 \tau^3= \left( \begin{array}{cc}
                          1 &0\\
                          0 &-1
                         \end{array} \right).
\]
Let us choose the loops to wind once around the generator
$\alpha_1$. As in the 2+1 dimensional case, we promote the loop
variables to operators on the Hilbert space of $L^2$ functions
$\Psi(a_1,a_2,a_3)$ over the reduced configuration space.  It then
follows that:
\beqa
\hat{T}^0[\alpha_1]\psi(a_1,a_2,a_3)=2\cos(a_1\pi)\psi(a_1,a_2,a_
3),  \n \\
\hat{T}^1[\hat{\alpha_1}] \Psi(a_1,a_2,a_3)
={1\over\pi^2}\int du^i\int ds^j\frac{1}{2}\ep_{ijk}
Tr[\pi^k U_{\alpha_1}(s)]\Psi(a_1,a_2,a_3).
\eeqa
Since on $\alpha_1$ the loop has just one component, i.e.
$ds^j=d\theta^1$,
we get
\beqa
\hat{T}^1[\hat{\alpha_1}] \Psi(a_1,a_2,a_3)= \n \\
\int_0^1 du\int_0^{2\pi} d\theta^1\frac{1}{2}
\{Tr[\pi^2 U_{\alpha_1}(s)]-Tr[\pi^3
U_{\alpha_1}(s)]\}\Psi(a_1,a_2,a_3)\n \\
=-{1\over\pi} i \hbar \sin(\pi a_1)[\frac{\partial}{\partial a_2}-
\frac{\partial}{\partial a_3}]\Psi(a_1,a_2,a_3,).
\eeqa
Generally, we thus have
\beqa
\hat{T}^0[\alpha_i]\Psi(a_1,a_2,a_3)=&2\cos(a_i\pi)\Psi(a_1,a_2,a
_3), \\
\hat{T}^1[\hat{\alpha_i}] \Psi(a_1,a_2,a_3)=&
-{1\over\pi} i \hbar \sin(\pi
a_i)\sum_{jk}\ep_{ijk}\frac{\partial}{\partial a_j}
\Psi(a_1,a_2,a_3,).
\eeqa
Since $\hat T^1$ is linear in the momenta, it can always be written
as
\[\hat{T}^1=\sum v^a (q) p_a,\]
at the point $(q,\;p)$ in the reduced phase space, where $v^\alpha
(q)$ is a vector
field on the configuration space. It is then the vector field
associated
with the operator $\hat{T}^1$ by:
\beqn
\hat{T}^1[\hat{\alpha}]\Psi(\omega)=
\frac{\hbar}{i}{\cal L}_{v_{\hat{\alpha}}}\Psi(\omega),
\eeqn
where ${\cal L}$ is the Lie derivative on ${\cal C}$. In the case of
$\Sigma=T^3$,
\[v_{\hat{\alpha}}^i=\int ds dt
\ep^{ijk}\delta^3(x,\,\hat{\alpha}(s,t))
\partial_s \alpha_j \partial_t \alpha_k
Tr(\tau^3U_{\hat{\alpha}})\;.\]

We now quantize in the loop representation.  The quantum theory is
given in terms of
a representation of the loop algebra as an operator
algebra \cite{ash:bo,smo:hop}.  The usual canonical quantization
consists
of a representation of the canonical Poisson algebra. By the gauge
property of the theory, we can construct a representation on
the gauge
constraint $C_{ab}=0$ surface of the phase space. The
representation
space ${\cal S}$ will consist of functionals of loops in $\Sigma$,
${\cal A}[{\alpha}]\in {\cal S}$.
The action of $\hat{T}^0[\alpha]$ and $\hat{T}^1 [\hat{\alpha}]$
can be
expressed in this representation as follows:
\beqa
&(\hat{T}^0[\alpha]{\cal A})[{\gamma}]:=
{\cal A}[\alpha\#{\gamma}]+{\cal A}[\alpha\#{\gamma^{-1}}],\\
&(\hat{T}^1[\hat{\alpha}]{\cal A})[\{\gamma\}]:=
i\hbar \sum_n I_n(\hat{\alpha}, \gamma)({\cal
A}[\hat{\alpha}\#_n\gamma]
-{\cal A}[\hat{\alpha}\#_n\gamma^{-1}]),
\eeqa
where we can use any loop $\gamma$ in the same homotopy class; the
result
is independent of the choice. This representation satisfies the
following
properties:
\beqas
&(1)&\;\;\hat{T}^0[\alpha]=\hat{T}^0[\alpha^{-1}],\;\;
   \hat{T}^1[\hat{\alpha}]=\hat{T}^1[\hat{\alpha}^{-1}],\\
&   &\hat{T}^0[\alpha\#\gamma]=\hat{T}^0[\gamma\#\alpha],\;\;
\hat{T}^1[\hat{\alpha}\#\hat{\gamma}]=\hat{T}^1[\hat{\gamma}\#\hat
{\alpha}];\\
&(2)&\;\;\hat{T}^0[0]=d\,,\;\;
\hat{T}^1[0]=0;\\
&(3)&\;\;\hat{T}^0[\alpha\#\gamma]+\hat{T}^0[\alpha\#\gamma^{-1}]
=\hat{T}^0[\alpha]\hat{T}^0[\gamma];\\
&(4)&\;\; \left[\hat{T}^A ,\;\hat{T}^B\right]=i\hbar
\{T^A,\;T^B\},\;\; A,\,B=0,\,1;\\
&(5)&\;\; \hat{T}^0[\alpha]=(\hat{T}^0[\alpha])^{*}\, is \, \;
real.
\eeqas
In (2) $d$ is the dimensions of the group representation.

In a given connected sector of the reduced phase space $\hat{\cal
C}$, denoted by
$\hat{\cal C}_i$, the transformation between the two sorts of
representations is \cite{ash:cqg}:
\beqn
{\cal A}[\alpha]=\int_{\hat{\cal C}_i}dV
(T^0[\alpha](\omega))\Psi(\omega),
\eeqn
where $dV$ is a volume element  on $\hat{\cal C}_i$ and
$\Psi(\omega)$ is the
wave function in the connection representation.

Returning to the case of the three torus discussed above, by
choosing
the volume element $dV=da_1 da_2 da_3$ on the torus, we could
construct
the transform explicitely in the time-like sector:
\beqn
{\cal A}[n_1, n_2, n_3]=\int_0^1da_1\int_0^1da_2\int_0^1da_3
\cos(\pi n_i a_i)\Psi(a_1, a_2, a_3)\,,\label{Trac}
\eeqn
where $n_i a_i=n_1 a_1+n_2 a_2+n_3 a_3$.

The fundamental operators $\hat{T}^0\,,\;\hat{T}^1$ in the loop
representation
are
\beqa
\hat{T}^0[\alpha_1]{\cal A}[n_1,n_2,n_3]=
{\cal A}[n_1+1, n_2,n_3]+{\cal A}[n_1-1, n_2, n_3]\label{t0n}\\
\hat{T}^1[\hat{\alpha}_1]{\cal A}[n_1,n_2,n_3]=\\ \n
i\hbar (n_2-n_3)\left({\cal A}[n_1+1, n_2,n_3]- {\cal
A}[n_1-1, n_2, n_3]\right)\label{t1n}
\eeqa
and similarly for the operators associated with the other two
homotopy group
generators $\alpha_2, \;\alpha_3$. Noticing that the transform
(\ref{Trac})
is just a cosine Fourier transform, we can represent the wave
functions
in the connection representation in terms of the wave functions in
the loop representation. The inverse transform is
\beqn
\Psi(a_1,a_2, a_3)={\cal A}[0,0,0]+\\ \n
2\sum_{n_1\,n_2\,n_3}
{\cal A}[n_1,n_2,n_3]\cos (\pi n_i a_i)\,.
\eeqn
This is a demonstration that the representation of the loop algebra
is (over) complete. Any wave function can be expanded as a
linear combination of the states ${\cal A}[n_1,n_2,n_3]$.

\section{Discussion}
Comparing the 4-D BF theory and Einstein gravity, we see that the
constraint
structure of these two theories are quite different. In terms of
Ashtekar's
variables, the constraints of Einstein gravity are \cite{ash:bo}:
\beqn
\begin{array}{l}
{\cal G}_a={\cal D}_i \tilde{E}^i_a\approx 0\,,\\
{\cal V}_i=\tilde{E}^j_aF^a_{ij}\approx 0\,,\\
{\cal S}=\ep^{abc} F_{ija}\tilde{E}^i_b\tilde{E}^j_c\approx 0\,.
\end{array}
\eeqn
They are usually called, respectively, the Gauss constraint, the
vector constraint and
the scalar (or Hamiltonian) constraint. The loop variables are
weakly annihilated by the Gauss
and the scalar constraints, but not by the vector constraint since
the latter generates diffeomorphisms. In contrast, in the 4-D BF
theory,
the vector constraint and the scalar constraint are equivalent to
the constraint
that the $SO(3,1)$ connection is flat. Thus there are only two
constraints,
\[
R_{ij}^{ab}\approx 0, \;\; C_{ab}\approx 0,
\]
in the BF theory. They generate the spatial projections of the
gauge
transformations. The gauge invariant loop variables naturally
commute with these constraints. Furthermore, in the BF theory
diffeomorphisms
are generated by the constraints \cite{ho:com}. In this sence, the
loop
variables {\it are} the physical observables.

Finally, let us note an interesting difference between the loop
representation
of the $T$-variables on $T^3$ in the 4-D BF theory and the loop
representation
of the $T$-variables on $T^2$ in the 3-D gravity theory
\cite{smo:hop,ash:cqg}.
Equations (\ref{t0n}) and (\ref{t1n}) tell us that on $T^3$, not
only is the
trivial state ${\cal A}[0,0,0]$ annihilated by
$\hat{T}^1$, but so are {\it all} the states ${\cal A}[n,n,n]$ for
any integer $n$.   Furthermore, these are the only states which
are annihilated by $\hat T^1$.
Hence, in the connection representation, the state $\Psi_0(\omega)$
with
$\hat{T}^1\Psi_0(\omega)=0$ can be written as
\beqn
\Psi_0(\omega)=\sum_n c_n(\omega){\cal A}[n,n,n]
\eeqn
where $n$ are arbitary integers.

We can also construct these observables in another simple
topological space, namely
 $\Sigma=S^2\times S^1$.  Since $\Sigma$ is now a product space,
the
fundamental group is a product \cite{hu:hom}:
\beqn
\pi_1(\Sigma)= \pi_1(S^2)\times\pi_1(S^1)\;.
\eeqn
But since $\pi_1(S^2)=\{e\}$,
\beqn
\pi_1(\Sigma)=\pi_1(S^1).
\eeqn
The homotopy group of $\Sigma$ has only one generator, denoted
$\alpha$.
Hence the homotopy of any single loop $\Gamma$ is labeled by the
integer $n$,
such that $\Gamma=\alpha^n$. The non-trivial holonomies are
generated
by a single homotopy class, and hence are determined by a single
axis
of given causality.  Let $(\theta,\, \theta^{\prime},\,\phi)$ be
the
coordinates on $\Sigma$, such that $(\theta,\,\theta^{\prime})$ are
the
coordinates of $S^2$. The ranges of the variables are then
\( 0\leq \theta\,,\phi< 2\pi\,,  0\leq\theta^{\prime}< \pi\,. \)
The configuration space $\hat{\cal C}$ is one dimensional.  For the
time-like
axis, the abelian subgroup of $SO(3,1)$ is $SO(2)$, and hence the
topology of
this sector is $S^1$. Introducing the coordinate $a\in [0,\;1]$ on
it, we may
choose a flat connection on $\Sigma$ representing that point of
$\hat{\cal C}$:
\beqn
\omega^b=\omega^b_i dx^i =a(\partial_i\phi)dx^i\tau^b=ad\phi
\tau^b,
\eeqn
where $\tau^b$ is the generator of $SO(2)$. If we choose the dual
of the
momenta as \[\pi^{*a}_{ij}=p
\partial_i\theta^A\partial_j\phi\tau^b,\]
with $n$ denoting the winding number of the loop, then the $T$
observables are
given by
\beqn
\hat{T}^0[n](a)=2\cos( n\pi a)\;\;\; , \;\;\;\hat{T}^1[n](a,p)=2\pi
p\sin( n\pi a).
\eeqn \\
\vspace{20 pt}
{\bf Acknowledgments} \\
The authors wish to acknowledge the partial support of the Natural
Sciences and Engineering Research Council of Canada.

\end{document}